\documentstyle[12pt,epsfig,amsfonts]{article}
\title{Renormalization and radiation reaction\\ in 2+1 electrodynamics}
\author{\bf Yurij Yaremko}
\date{\it Institute for Condensed Matter Physics, \\
1 Svientsitskii St., 79011 Lviv, Ukraine}
\begin{document}
\maketitle
\begin{abstract}
We consider a self-action problem for an electric charge arbitrarily moving 
in flat spacetime of three dimensions. Its electromagnetic field satisfies 
the Maxwell equations in Minkowski space ${\mathbb M}_{\,3}$ of three 
dimensions. In ${\mathbb M}_{\,3}$ electromagnetic waves propagate not just 
at a speed of light, but also at all speeds smaller than or equal to the 
speed of light. The massive particle may ``fill'' its own field, which acts 
on it just like an external one. The radiation reaction is determined by 
Lorentz force of point-like charge acting upon itself plus non-local term 
which provides finiteness of the self-action. The self-force produces a 
time-changing inertial mass. The relation between $2+1$-electrodynamics and 
dynamics of superfluid $\!\!\!\phantom{I}^4$He film is emphasized.
\end{abstract}




\section{Introduction}\label{intro}
\setcounter{equation}{0}

There is an extensive literature devoted to the physics of a superfluid 
$\!\!\!\phantom{1}^4{\rm He}$. Electrons on the surface of liquid helium are 
a widely studied system promising implementation of quantum computer 
\cite{SSL}. Adsorption of thin films of $\!\!\!\phantom{1}^4{\rm He}$ on 
various substrates is motivated by the intriguing properties of these films
\cite{Bon}. In Refs.\cite{LT,TL} the influence electromagnetic field induced 
in a dielectric disk resonator placed in He-II on the dynamics of phonon and 
roton excitations is considered. (Since pioneer works by L.Landau 
\cite{LL1,LL2}, $\!\!\!\phantom{1}^4{\rm He}$ is a quantum fluid in which 
motions exist in the form of quasiparticles.) In superfluid helium occur 
also quantized vortices \cite{Don} which play a key role in the dissipation 
of energy and momentum. Vortices represent the breakdown of laminar fluid 
flow. The fluid rotation associated with a vortex can be parametrized by the 
circulation $q=\oint {\rm d}{\mathbf l}\cdot {\mathbf v}({\mathbf l})$ 
about the vortex, where ${\mathbf v}({\mathbf l})$ is the fluid velocity 
field. While classical vortices can take any value of circulation, in a 
superfluid film the rotation occurs through vortices with quantized 
circulation. 

There exists the remarkable correspondence between dynamical equations 
which govern behavior of superfluid $\!\!\!\phantom{I}^4$He films and 
Maxwell equations for electrodynamics in $2+1$ dimensions \cite{AHNS}. 
It is of great importance that the dynamics of the low energy 
quasiparticles and elementary excitations living inside a helium film is 
governed by Maxwell equations in ${\Bbb M}_3$. Therefore, if one study the 
behavior of electric charges living inside hypothetic spacetime with two 
space directions, they study the kinetic of vortices and phonon excitations 
in superfluid ${\rm He}^4$ film. In the present paper we establish the 
equation of motion of a point electric charge in ${\Bbb M}_3$ under the 
influence of an external electromagnetic field, where the effects of 
radiation reaction are taken into account.

The computation of effect of particle's own field is not a trivial matter, 
since the Green's function associated with the wave operator has support 
within the light cone. This is because in three dimensions electromagnetic 
waves propagate not just at a speed of light, but also at all speeds smaller 
than or equal to the speed of light. The particle may ``fill'' its own 
field, which acts on it just like an external one. The equation of motion 
require one to identify that portion of the retarded field at each point of 
the world line which arises from source contributions interior to the 
light cone. This part of field is often called the ``tail term''. The self 
force on a particle then consists of two parts: this comes from the direct 
part of the Green's function and depends on the current state of 
particle's motion and that comes from the tail part and depends not only 
the current state of the particle, but also on its past history. It leads 
to the non-local (integro-differential) equations of motion. 

Detweiler and Whiting proposed \cite{DW} a consistent decomposition of the 
retarded Green's function into singular and radiative parts. It obeys the 
spirit of Dirac's scheme of splitting of electromagnetic potential of a 
point-like charged particle arbitrarily moving in flat spacetime. Dirac 
\cite{Dir} decomposed the retarded Li\'enard-Wiechert potential $A^{\rm 
ret}$ into two parts: (i) one-half of the retarded plus one-half of the 
advanced potentials which is inhomogeneous solution of the wave equation 
$\square A_\alpha=-4\pi j_\alpha$ whose source term is infinite on the 
world line. $A^{\rm S}=1/2(A^{\rm ret}+A^{\rm adv})$ is just singular as 
the retarded potential in the immediate vicinity of the particle's world 
line. The superscript ``S'' stands for ``singular'' as well as 
``symmetric''. (ii) combination $A^{\rm R}=1/2(A^{\rm ret}-A^{\rm adv})$ of 
one-half of the retarded minus one-half of the advanced potentials which 
satisfies the homogeneous wave equation. This well behaved potential can be 
thought as a free radiation field. The superscript ``R'' stands for 
``radiative'' as well as ``regular''.

The radiative Green's function implicitly used by Dirac in flat spacetime 
is
\begin{equation}\label{Grad}
G^{\rm rad}(x,y)=G^{\rm ret}(x,y)-G^{\rm sym}(x,y)
\end{equation}
where
\begin{equation}\label{Gsym}
G^{\rm sym}(x,y)=\frac12\left[G^{\rm ret}(x,y)+G^{\rm adv}(x,y)\right].
\end{equation}
The causal structure of the Green's function is richer in curved 
spacetime. Due to contributions of the interior of the light cones, the 
retarded potential depends on the particle's history {\it prior} to the 
retarded instant $\tau^{\rm ret}(x)$ while the advanced one is generated 
by portion of particle's world line $\zeta$ {\it after} the advanced 
instant $\tau^{\rm adv}(x)$. (The retarded and the advanced moments label 
the points on $\zeta$ related with arbitrary field point $x\in{\mathbb 
M}_{\,4}$ by null rays.) While the combination of half-retarded minus 
half-advanced potential would satisfy the homogeneous wave equation and it 
would be smooth on the world line, a self-force constructed from this 
radiative potential would be highly non-causal. It would be depend on 
particle's entire history, both past (through the retarded Green's 
function) and future (through the advanced Green's function). The Dirac's 
scheme (\ref{Grad}) for decomposition cannot be adopted without 
modification in curved spacetime. The modification is performed in 
Ref.\cite{DW}. Following their scheme, Detweiler and Whiting recovered the 
results \cite{WB}-\cite{Q} for electromagnetic, scalar, and gravitational 
fields.

It is obvious that the physically relevant solution of the wave equation 
is the retarded potential. Teitelboim \cite{Teit} derived the 
electromagnetic self-force in flat spacetime within the framework of 
retarded causality. The author substituted the retarded Li\'enard-Wiechert 
field in the Maxwell energy-momentum tensor density and calculated the 
flow of energy-momentum which flows across a space-like surface. Minkowski 
space was parameterized by four curvilinear coordinates. The first, proper 
time, labels points of emission placed on $\zeta$, the second one 
determines the surface (e.g., a tilted hyperplane which is orthogonal to 
particle's 4-velocity at fixed instant of observation). Having integrated 
the stress-energy tensor over two angular variables that distinguish 
points on the surface, Teitelboim found the flow of energy-momentum 
mentioned above. The resulted expression depends on the particle's 
individual characteristics (on its mass, its charge, its velocity and 
acceleration). In fact, the surface integration is equivalent to taking of 
coincidence limit in Dirac's scheme. Abraham-Lorentz-Dirac expression for 
electromagnetic self-force is obtained in \cite{Teit} via consideration of 
energy-momentum conservation. 
 
In previous papers \cite{Yar3D,Y3D} we have used Teitelboim's approach to 
take proper account of contribution from interior of the light cone. For 
the clearest demonstration of the impact of our analysis, we have 
considered a point-like particle of mass $m$ and charge $e$ coupled to 
electromagnetic field in flat spacetime of three dimensions. In the present 
paper we summarize the consideration \cite{Yar3D,Y3D} as a consisytent 
regularization procedure which exploits the Poincar\'e invariance of the 
theory. 

\section{Maxwell equations in ${\mathbb M}_{\,3}$}
\setcounter{equation}{0}

We consider an electromagnetic field produced by current ${\mathbf j}=
j^0{\mathbf e}_0+j^1{\mathbf e}_1+j^2{\mathbf e}_2$. In terms of 
differential forms the Maxwell equations look as usual:
\begin{eqnarray}\label{Maxw} 
{\rm d}{\hat F}&=&0,\\ {\rm d}^{\,*}\!{\hat F}&=&2\pi^{\,*}\!{\hat j}. 
\label{MaxwD}
\end{eqnarray} 
In three dimensions the components of Faraday 2-form are
\begin{equation}\label{Far2}
(F_{\alpha\beta})=\left(
\begin{array}{ccc}
0&-E^1&-E^2\\[1ex]
E^1&0&H\\[1ex]
E^2&-H&0
\end{array}\right).
\end{equation}
The electric field $E^i$ has two components while the magnetic field $H$ 
has only one component. Under a spatial rotation $F_{\alpha\beta}$ 
transforms in such a way that $(E^1,E^2)$ transforms as two-vector while 
$H$ does not change at all.

Exterior derivative of the Faraday two-form (\ref{Far2}) is 3-form 
\begin{equation}\label{dFar}
{\rm d}\hat F=\left(\frac{\partial H}{\partial x^0}-\frac{\partial 
E^1}{\partial x^2}
+\frac{\partial 
E^2}{\partial x^1}\right){\rm d}x^0\wedge{\rm d}x^1\wedge{\rm d}x^2.
\end{equation}
Since ${\rm d}\hat F=0$, the bracketed expression vanishes:
\begin{equation}\label{FarL}
\frac{\partial H}{\partial x^0}-\frac{\partial E^1}{\partial x^2}
+\frac{\partial 
E^2}{\partial x^1}=0.
\end{equation}
It is the Faraday's law of induction in $2+1$ electrodynamics.

By means of metric tensor $\eta_{\alpha\beta}={\rm diag}(-1,1,1)$, we 
define an antisymmetric ${2\choose 0}$ tensor $\mathbf F$, whose 
components are
\begin{eqnarray}\label{MaxFD}
F^{\mu\nu}&=&\eta^{\mu\alpha}\eta^{\nu\beta}F_{\alpha\beta},\\
(F^{\mu\nu})&=&\left(
\begin{array}{ccc}
0&E^1&E^2\\[1ex]
-E^1&0&H\\[1ex]
-E^2&-H&0
\end{array}\right).\nonumber
\end{eqnarray}
Because $({\bf e}_0,{\bf e}_1,{\bf e}_2)$ form an orthonormal basis in 
flat spacetime metric $\eta_{\alpha\beta}$, the volume three-form in 
${\mathbb M}_{\,3}$ is
$$
{\hat\omega}={\rm d}x^0\wedge{\rm d}x^1\wedge{\rm d}x^2. 
$$
1-form $\!\!\phantom{1}^{\,*}\!{\hat F}$ in eq.(\ref{MaxwD}) is the 
Hodge dual of tensor (\ref{Far2}):
\begin{eqnarray}
\!\!\phantom{1}^{\,*}\!{\hat F}&=&\frac12{\hat\omega}({\mathbf F})\nonumber\\
&=&H{\rm d}x^0-E^2{\rm d}x^1+E^1{\rm d}x^2.\nonumber
\end{eqnarray}
Magnetic field $H$ is its zeroth component. Exterior derivative 
\begin{eqnarray}\label{d*F}
{\rm d}^{\,*}\!{\hat F}&=&
\left(\frac{\partial E^1}{\partial x^1}+\frac{\partial E^2}{\partial 
x^2}\right){\rm d}x^1\wedge{\rm d}x^2+
\left(\frac{\partial E^1}{\partial x^0}
-\frac{\partial H}{\partial x^2}
\right){\rm d}x^0\wedge{\rm d}x^2\\
&-&
\left(
\frac{\partial E^2}{\partial x^0}+\frac{\partial H}{\partial x^1}
\right){\rm d}x^0\wedge{\rm d}x^1\nonumber
\end{eqnarray}
is equal to dual current 
\begin{eqnarray}\label{d*j}
\!\!\phantom{1}^{\,*}\!{\hat j}&=&{\hat\omega}({\mathbf j})\\
&=&j^0{\rm d}x^1\wedge{\rm d}x^2
-j^1{\rm d}x^0\wedge{\rm d}x^2 + j^2{\rm d}x^0\wedge{\rm d}x^1.\nonumber
\end{eqnarray}
Comparing the components in the right-hand sides of eq.(\ref{d*F}) and 
eq. (\ref{d*j}), we obtain the system of differential equations in partial 
derivatives:
\begin{equation}\label{GsAmp}
\frac{\partial E^1}{\partial x^1}+\frac{\partial E^2}{\partial x^2}=2\pi 
j^0,\qquad
-\frac{\partial E^1}{\partial x^0}+\frac{\partial H}{\partial x^2}=2\pi 
j^1,\qquad
-\frac{\partial E^2}{\partial x^0}-\frac{\partial H}{\partial x^1}=2\pi 
j^2.
\end{equation}

These expressions together with eq.(\ref{FarL}) are the Maxwell equations 
(\ref{Maxw}) and (\ref{MaxwD}) in $2+1$ electrodynamics. In the next 
Section we review a well-known similarity between physics of a superfluid 
film and classical electrodynamics in Minkowski space of three dimensions 
\cite{AHNS}. We show that the dynamics of the low energy quasiparticles and 
elementary excitations living inside a helium film is governed by Maxwell 
equations (\ref{FarL}) and (\ref{GsAmp}).

\section{Superfluid He-II film as $2+1$-electrodynamics}\label{Helium}
\setcounter{equation}{0}

In the superfluid state macroscopic numbers of helium atoms are in ground 
state with zero momentum. What is the form of the wave function of this 
condensate of identical bosons? If the condensate is static and 
homogeneous, the wave function is constant:
$$
\psi(t,{\mathbf x})=\sqrt{n_0}.
$$
($n_0$ is the numbers of the ground state particles at unite volume.) 
Whenever the system is unstable and inhomogeneous, its wave function 
$$
\psi(t,{\mathbf x})=\sqrt{n_0(t,{\mathbf x})}\exp\left[i\phi(t,{\mathbf 
x})\right]
$$
has a well-defined $U(1)$ order parameter $\phi(t,{\mathbf x})$ which is 
called a {\it phase}. The first Josephson equation of superfluidity 
\begin{equation}\label{Js1}
v_i(t,{\mathbf x})=\frac{\hbar}{m}\frac{\partial\phi(t,{\mathbf 
x})}{\partial x^i}
\end{equation}
introduces the velocity field $(v_1(t,{\mathbf x}),v_2(t,{\mathbf x}))$ 
that describes time evolution of excitations in a helium film. These 
excitations are meant as quasiparticles of two kinds: sound waves (phonons) 
and vortices. An investigator measures the velocity $v_i(t,{\mathbf x})$ 
and density $\rho(t,{\mathbf x})$ fields of phonons in presence of vortices 
with their own densities $\rho_v(t,{\mathbf x})$ and currents 
${\mathbf j}_v=(j^1_v(t,{\mathbf x}),j^2_v(t,{\mathbf x}))$. Phonon 
parameters satisfy the equation of continuity
$$
\frac{\partial \rho}{\partial t}+\frac{\hbar}{m}\frac{\partial{{\bar\rho} 
v_i(t,{\mathbf x})})}{\partial x^i}=0,
$$
where ${\bar\rho}$ is average density of the fluid. If we make the following 
identification
\begin{equation}\label{HeEd}
H\Leftrightarrow -c\frac{\rho}{\bar\rho}\,,\quad
E^1\Leftrightarrow -v^2 \,,\quad E^2\Leftrightarrow v^1
\end{equation}
we see that the equation of continuity becomes the Faraday's law of induction
(\ref{FarL}). 

Sound waves (phonon excitations) map onto the electromagnetic fields, 
while vortices map onto electric charges. A vortex can be thought as a 
circular
flow of the fluid around a core of a very small radius. (Two kinds of 
charges correspond to clockwise and counterclockwise directions of motion.)
The flow around a core is quantized:
\begin{equation}\label{intHe}
\oint ({\mathbf v}\cdot{\rm d}{\mathbf l})=2\pi\frac{\hbar}{m}q,
\end{equation} 
where $m$ is mass of $He^4$ atom. Quantized parameter $q$ is called a {\it 
vorticity}\footnote{The vortices of minimal value $q\pm 1$ only are meant 
to be thermodynamically stable.}. The integral equation (\ref{intHe}) can be 
rewritten in form of equivalent differential equation
$$
-\frac{\partial v_2}{\partial x^1}+\frac{\partial v_1}{\partial x^2}=
2\pi\rho_v,
$$
where $\rho_v(t,{\mathbf x})$ is $\delta$-shaped density of vortex. If we 
identify the vortex density with the electric charge density (\ref{je3}), we 
arrive at the Gauss's law in three dimensions:
$$
\frac{\partial E^1}{\partial x^1}+\frac{\partial E^2}{\partial x^2}=
2\pi\rho_v.
$$

\begin{table}[]
\begin{center}
\caption{Correspondence between dynamics of ${\rm He}^4$ film and 
classical electrodynamics in three dimensions}
\vspace*{0.01\textheight}
\begin{tabular}{|p{0.45\textwidth}|p{0.49\textwidth}|}
\hline \hline
Superfluid ${\rm He}^4$ film& $2+1$ electrodynamics\\
\hline \hline
Phonon velocity field $(v^1,v^2)$ & Electric field $(E^1,E^2)$\\
Phonon density $\rho$& Magnetic field $H$\\
Vorticity $q\;(\pm 1)$& Electric charge $e$\\
Density of vortices $\rho_v$& Density of charges $\rho$\\
Current of vortices $(j^1_v,j^2_v)$& Electric current $(j^1,j^2)$\\
Parameter $\sqrt{\kappa/m}$ & Speed of light $c$\\
\hline
Equation of continuity 
$$\frac{\partial \rho}{\partial t}+\frac{\hbar}{m}\frac{\partial{{\bar\rho} 
v_i(t,{\mathbf x})})}{\partial x^i}=0$$
& Faraday's law of induction 
$$\frac{\partial H}{\partial x^0}-\frac{\partial E^1}{\partial x^2}
+\frac{\partial 
E^2}{\partial x^1}=0$$
\\[-1.5em]
\hline
Condition of quantization of vorticity 
$$
-\frac{\partial v_2}{\partial x^1}+\frac{\partial v_1}{\partial x^2}=
2\pi\rho_v
$$
& Gauss's law
$$
\frac{\partial E^1}{\partial x^1}+\frac{\partial E^2}{\partial x^2}=2\pi 
j^0
$$
\\[-1.5em]
\hline
Josephson laws of superfluidity 
$$
\frac{\partial v_i}{\partial t}+\frac{\kappa}{m{\bar\rho}}\frac{\partial 
\rho}{\partial x^i}=2\pi j_v^i
$$
& Amper's law
$$
-\frac{\partial E^1}{\partial x^0}+\frac{\partial H}{\partial x^2}=2\pi 
j^1,\;
-\frac{\partial E^2}{\partial x^0}-\frac{\partial H}{\partial x^1}=2\pi 
j^2$$
\\
\hline
\end{tabular}
\end{center}
\label{tabHe}
\end{table}

And, finally, Amper's law
\begin{equation}\label{Amp}
-\frac{\partial E^1}{\partial x^0}+\frac{\partial H}{\partial x^2}=2\pi 
j^1,\qquad
-\frac{\partial E^2}{\partial x^0}-\frac{\partial H}{\partial x^1}=2\pi 
j^2,
\end{equation}
corresponds to the combination of the Josephson equations of superfluidity.
The second Josephson equation of superfluidity 
\begin{equation}\label{Js2}
\hbar\frac{\partial\phi(t,{\mathbf x})}{\partial t}=-\mu(x)
\end{equation}
relates the time derivative of the phase to the chemical potential $\mu(x)$.
Differentiating the first equation (\ref{Js1}) with respect to time and 
taking into account eq.(\ref{Js2}), we obtain the change of velocity caused 
by inhomogeneity:
\begin{eqnarray}\label{Js12}
\frac{\partial v_i(t,{\mathbf x})}{\partial 
t}&=&-\frac{1}{m}\frac{\partial\mu(x)}{\partial x^i}\\
&=&-\frac{\kappa}{m}\frac{\partial\rho(x)}{\partial x^i}.\nonumber
\end{eqnarray}
We use the compressibility
$$
\kappa={\bar\rho}\frac{\partial\mu}{\partial\rho}
$$
to express the chemical potential $\mu$ in terms of the density $\rho$. 
In presence of the vortex flow which satisfies the equation of continuity
\begin{equation}\label{jcont}
\frac{\partial\rho_v}{\partial t}+\frac{\partial j_v^1}{\partial x^1}
+\frac{\partial j_v^2}{\partial x^2}=0,
\end{equation}
the equation (\ref{Js12}) modifies as follows:
$$
\frac{\partial v_i}{\partial t}+\frac{\kappa}{m{\bar\rho}}\frac{\partial 
\rho}{\partial x^i}=2\pi j_v^i.
$$
If we apply the rule (\ref{HeEd}) and identify the speed of light as 
$c^2=\kappa/m$, we arrive exactly at the Amper's law (\ref{Amp}).

The above considerations can be summarized in table 1.

\section{Electromagnetic potentials in 2+1 theory}\label{potn3}
\setcounter{equation}{0}

Since eq.(\ref{Maxw}), Faraday 2-form $\hat F$ is closed. Whence there is 
1-form $\hat A$ such that $\hat F={\rm d}\hat A$ in some neighbourhood of 
any point $x\in {\mathbb M}_{\,3}$. $\hat A$ is called one-form potential.
In Cartesian coordinates their components are related as follows:
$$
F_{\alpha\beta}=\frac{\partial A_\beta}{\partial x^\alpha}
-\frac{\partial A_\alpha}{\partial x^\beta}.
$$
Substituting this into eq.(\ref{MaxwD}) give rise to the second order 
differential equation for the one-form potential with a charge density 
$j_\mu(x)$:
$$
\square A_\mu=-2\pi j_\mu,
$$
where $\square=\eta^{\alpha\beta}\partial_\alpha\partial_\beta$ is 
D'Alembert differential operator. (The Lorentz gauge is imposed.)

In $2+1$ electrodynamics the retarded Green's function associated with 
D'Alembert operator is supported within the light cone \cite{Gl,KLS}:
\begin{equation}\label{G3ret}
G^{\rm ret}_{2+1}(x,y)=\frac{\theta(x^0-y^0-|{\mathbf x}-{\mathbf 
y}|)}{\sqrt{-2\sigma(x,y)}}.
\end{equation}
$\theta(x^0-y^0-|{\mathbf x}-{\mathbf y}|)$ is the light cone step 
function defined to be one if $x^0-y^0\ge|{\mathbf x}-{\mathbf y}|$ and 
defined to be zero otherwise. Synge's world function $\sigma(x,y)$ is 
numerically equal to half the squared distance between $x$ and $y$. The 
analysis of the simplest model with tails will give more deep understanding 
of Detweiler and Whiting scheme of decomposition. 

We consider an electromagnetic potential produced by a particle with 
$\delta$-shaped distribution of the electric charge $e$ 
\begin{equation}\label{je3}
j^\alpha=e\int_{-\infty}^{+\infty}{\rm d}\tau 
u^\alpha (\tau)\delta^{(3)}(y-z(\tau)),
\end{equation} 
moving on a world line $\zeta\subset {\mathbb M}_{\,3}$ described by 
functions $z^\mu(\tau)$ of proper time $\tau$.

\begin{figure}[ht]
\begin{center}
\epsfclipon
\epsfig{file=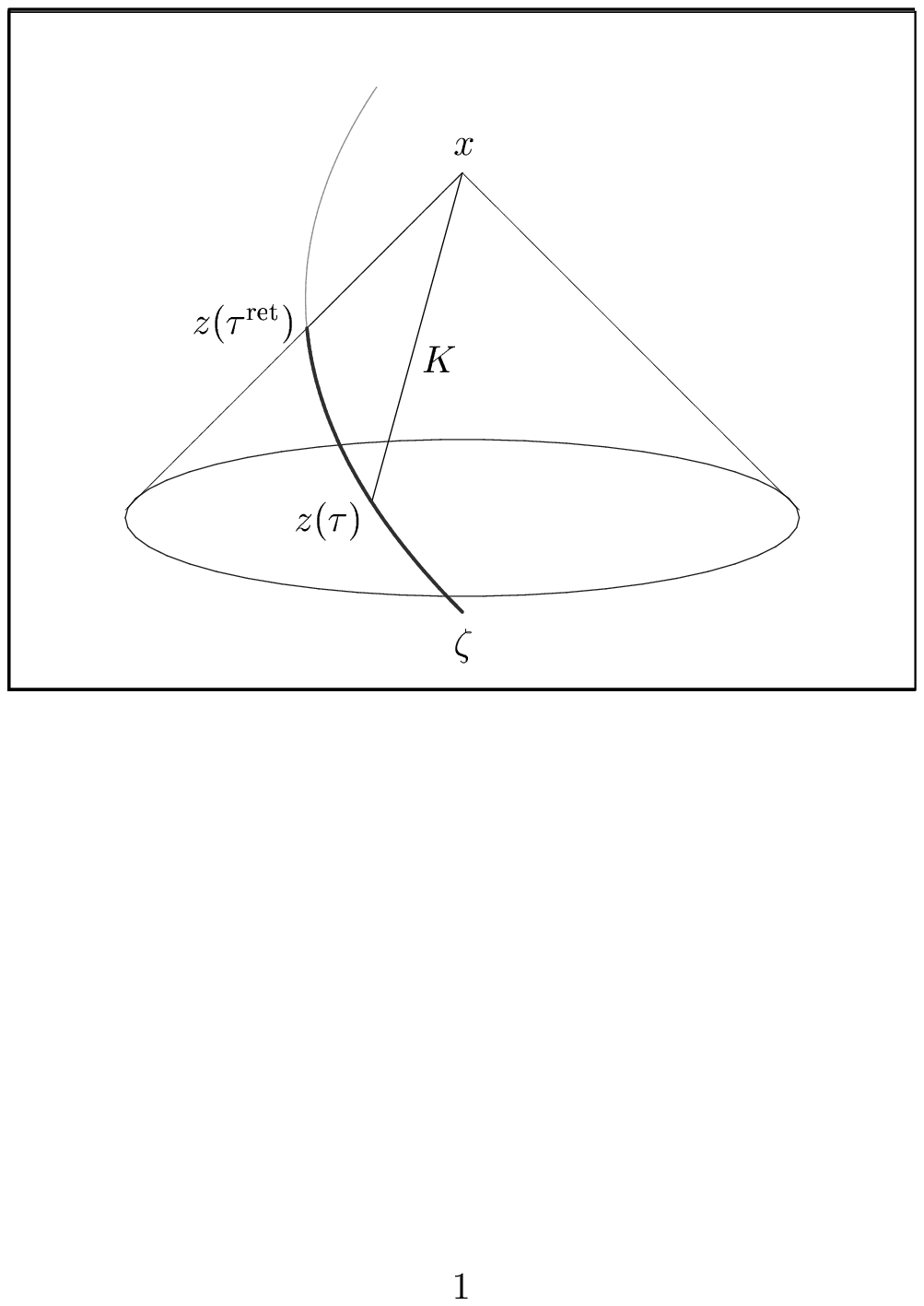,width=0.45\textwidth}
\end{center}
\caption{In four dimensions the retarded potential at field point $x$ is 
generated by a single event in space-time: the intersection $z(\tau^{\rm 
ret})$ of the world line $\zeta$ and $x$'s past light cone. In three 
dimensions the retarded potential depends on the segment of $\zeta$ that 
corresponds to time interval $]-\infty,\tau^{\rm ret}(x)]$. The vector $K$ 
is a vector pointing from the emission point $z(\tau)\in\zeta$ to field 
point $x$.
}\label{ret1}
\end{figure}

Convolving the retarded Green function (\ref{G3ret}) with charge-current 
density (\ref{je3}), we construct the retarded Li\'enard-Wiechert potential 
in three dimensions: 
\begin{equation} \label{Aret3}
A_\mu^{\rm ret}(x)=e\int_{-\infty}^{\tau^{\rm ret}(x)}{\rm d}\tau 
\frac{u_\mu(\tau)}{\sqrt{-(K\cdot K)}}.
\end{equation}
It is generated by the point charge during its entire past history before 
the retarded time $\tau^{\rm ret}(x)$ associated with the field point $x$
(see figure \ref{ret1}). We denote $K^\mu=x^\mu - z^\mu(\tau)$ the unique 
timelike (or null) vector connecting a field point $x$ to the emission 
point $z(\tau)\in\zeta$. The dot denotes the scalar product of three-vector 
$K$ on itself. $(K\cdot K)$ is equal to double Synge's function 
$\sigma(x,z(\tau))$ of field point $x$ and emission point $z(\tau)$.

The simplest potential is produced by an unmoved charge placed at the 
coordinate origin. The world line $\zeta$ is given by $z^\mu=(t,0,0)$. The 
retarded instant $\tau^{\rm ret}(x)=x^0-r$ where $r:=\sqrt{(x^1)^2+(x^2)^2}$ 
is distance from the field point with coordinates $(x^0,x^1,x^2)$ to the 
charge. Since 3-velocity $u_\mu=(-1,0,0)$, the only nontrivial components of 
the retarded potential (\ref{Aret3}) is
\begin{eqnarray}
A_0^{\rm ret}&=&e\int_{-\infty}^{x^0-r}{\rm d}t 
\frac{-1}{\sqrt{(x^0-t)^2-r^2}}\nonumber\\
&=&\left.-e\ln\left(x^0-t-\sqrt{(x^0-t)^2-r^2}\right)
\right|_{t\to -\infty}^{t=x^0-r}\nonumber\\
&=&-e\ln r.\nonumber
\end{eqnarray}
Having used metric tensor $\eta^{\alpha\beta}={\rm diag}(-1,1,1)$ to raise 
index, we arrive at the logarithmic static potential 
$$
A^0_{\rm ret}=e\ln r.
$$ 

\begin{figure}[ht]
\begin{center}
\epsfclipon
\epsfig{file=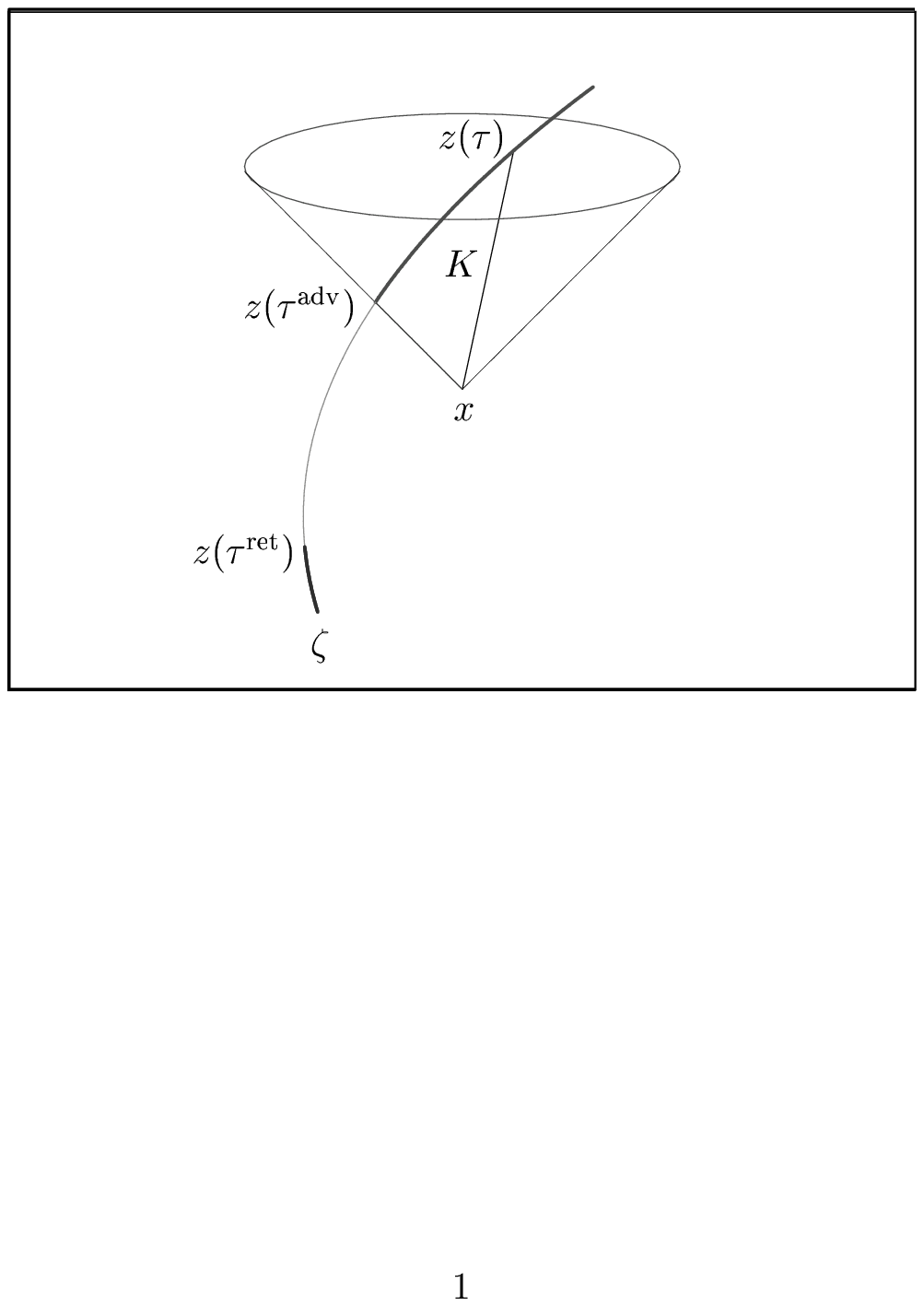,width=0.4\textwidth}
\end{center}
\caption{The portion of the world line after $\tau^{\rm adv}(x)$ produces 
the advanced potential in three dimensions. Advanced instant labels point 
at which particle's world line punctures future light cone with vertex at 
field point $x$. Retarded instant $\tau^{\rm ret}(x)$ associated with point 
$x$ is before the moment $\tau^{\rm adv}(x)$.
}\label{adv1}
\end{figure}

The advanced Green's function 
$$
G_{2+1}^{\rm adv}(x,y)=\frac{\theta(-x^0+y^0+|{\mathbf x}-{\mathbf 
y}|)}{\sqrt{-2\sigma(x,y)}}.
$$
is nonzero in the past of the emission point $y$. Convolving it with 
charge-current density (\ref{je3}), we construct the advanced potential:
$$
A_\mu^{\rm adv}(x)=e\int^{+\infty}_{\tau^{\rm adv}(x)}{\rm d}\tau 
\frac{u_\mu(\tau)}{\sqrt{-(K\cdot K)}}.
$$
It is generated by the point charge during its entire future history 
following the advanced time $\tau^{\rm adv}(x)$ (see figure \ref{adv1}).

Static advanced potential coincides with the retarded one.

\section{Electromagnetic field in 2+1 electrodynamics}\label{field3}
\setcounter{equation}{0}

The field $F_{\mu\nu}^{\rm ret}=\partial_\mu A^{\rm ret}_\nu -\partial_\nu 
A^{\rm ret}_\mu$ consists of two quite different terms. The first 
term is due to dependence of the upper limit in path integral (\ref{Aret3}) 
on field point $x$:
\begin{equation}\label{Fd}
F_{\mu\nu}^{(\delta)}=\lim_{\tau\to\tau^{\rm ret}}\frac{e}{\sqrt{-(K\cdot 
K)}}\frac{u_\mu K_\nu - u_\nu K_\mu}{-(K\cdot u)}.
\end{equation}
We take into account that 
$$
\frac{\partial\tau^{\rm ret}(x)}{\partial x^\mu}=-k_\mu,
$$
where $k_\mu=K_\mu/r, r=-(K\cdot u)$. Since $\tau^{\rm ret}(x)$ is the root 
of algebraic equation 
\begin{eqnarray}
(K\cdot K)&=&
\eta_{\alpha\beta}(x^\alpha-z^\alpha(\tau))(x^\beta-z^\beta(\tau))
\nonumber\\
&=&0,\nonumber
\end{eqnarray}
direct term ${\hat F}^{(\delta)}$ diverges. 

The second term 
\begin{equation}\label{Fth}
F_{\mu\nu}^{(\theta)}=-e\int_{-\infty}^{\tau^{\rm ret}(x)}d\tau 
\frac{u_\mu K_\nu - u_\nu K_\mu}{[-(K\cdot K)]^{3/2}}
\end{equation}
arises from differentiation of expression under the integral sign in 
eq.(\ref{Aret3}).

The direct part (\ref{Fd}), which depends on the momentary characteristics 
of the charge, and the tail part (\ref{Fth}), which depends on 
its previous evolution, separately diverge on the light cone. The 
singularity, however, can be removed from the sum of ${\hat F}^{(\delta)}$ 
and ${\hat F}^{(\theta)}$. Using the identity
$$
\frac{1}{[-(K\cdot K)]^{3/2}}=\frac{1}{-(K\cdot 
u)}\frac{\rm d}{{\rm d}\tau}\frac{1}{\sqrt{-(K\cdot K)}}
$$
in eq.(\ref{Fth}) yields
\begin{eqnarray}\label{Ft}
F_{\mu\nu}^{(\theta)}&=&-\frac{e}{\sqrt{-(K\cdot K)}}\left.
\frac{u_\mu K_\nu - u_\nu K_\mu}{-(K\cdot u)}
\right|_{\tau\to -\infty}^{\tau\to\tau^{\rm ret}(x)}
\\
&+&e\int_{-\infty}^{\tau^{\rm ret}(x)}
\frac{{\rm d}\tau}{\sqrt{-(K\cdot K)}}\left\{
\frac{u_\mu K_\nu - u_\nu K_\mu}{[-(K\cdot u)]^2}\left[1+(K\cdot a)\right]
+\frac{a_\mu K_\nu - a_\nu K_\mu}{-(K\cdot u)}
\right\}\nonumber
\end{eqnarray}
after integration by parts. Summing up (\ref{Fd}) and (\ref{Ft}) and 
taking into account that $1/\sqrt{-(K\cdot K)}$ vanishes whenever 
$\tau\to -\infty$\footnote{We assume that average velocities are not 
large enough to initiate particle creation and annihilation, so that 
``space contribution'' $|{\mathbf K}|$ can not match with an extremely 
large zeroth component $K^0$.}, we finally obtain the expression
\begin{equation}\label{Fret3}
{\hat F}^{\rm ret}(x)= e\int_{-\infty}^{\tau^{\rm ret}(x)}
\frac{{\rm d}\tau }{\sqrt{-(K\cdot K)}}\left\{
\frac{u\wedge K}{r^2}\left[1+(K\cdot a)\right]+
\frac{a\wedge K}{r}
\right\},
\end{equation}
which is regular on the light cone. (It diverges on the particle's 
trajectory only where field point $x$ approaches emission point 
$z\in\zeta$.) Symbol $\wedge$ denotes the wedge product. The invariant 
quantity
\begin{eqnarray}\label{rint}
r&=&-(K\cdot u)\\
&=&-\eta_{\alpha\beta}(x^\alpha-z^\alpha(\tau))u^\beta(\tau)\nonumber
\end{eqnarray}
is an affine parameter on the time-like (null) geodesic that links $x$ to 
$z(\tau)$; it can be loosely interpreted as the time delay between $x$ and
$z(\tau)$ as measured by an observer moving with the particle.
Because the speed of light is set to unity, parameter $r(\tau^{\rm ret})$ is 
also the spatial distance between $z(\tau^{\rm ret})$ and $x$ as measured in 
this momentarily comoving Lorentz frame.

Calculation of the advanced strength tensor
$$
{\hat F}^{\rm adv}(x)= e\int^{+\infty}_{\tau^{\rm adv}(x)}
\frac{{\rm d}\tau }{\sqrt{-(K\cdot K)}}\left\{
\frac{a\wedge K}{r}
+\frac{u\wedge K}{r^2}\left[1+(K\cdot a)\right]
\right\}
$$
is identical to that of the retarded one and we do not bother with details. 
Advanced field is generated by the point charge during its entire future 
history following the advanced time associated with $x$ (see figure 
\ref{adv1}).

\section{Field of a uniformly moving charge}
\setcounter{equation}{0}
To calculate the electromagnetic field generated by a uniformly moving 
charge we substitute the functions which parameterize the straightforward 
world line
$$
z^\mu(\tau)=z^\mu_0+u^\mu\tau
$$
in eq.(\ref{Fret3}). We split the vector $K^\mu=x^\mu-z^\mu(\tau)$ into 
constant vector $K^\mu_0=x^\mu-z^\mu_0$ and time-dependent part:
$$
K^\mu=K^\mu_0-u^\mu\tau.
$$
The square of this 3-vector is quadratic in proper time $\tau$:
$$
-(K\cdot K)=\tau^2+2(K_0\cdot u)\tau -(K_0\cdot K_0).
$$
The retarded instant
$$
\tau^{\rm ret}(x)=-(K_0\cdot u)-\sqrt{(K_0\cdot u)^2+(K_0\cdot K_0)}
$$
and the advanced one
$$
\tau^{\rm adv}(x)=-(K_0\cdot u)+\sqrt{(K_0\cdot u)^2+(K_0\cdot K_0)}
$$
satisfy the light cone equation $(K\cdot K)=0$.

Since particle's acceleration vanishes and the wedge product 
$u\wedge K=u\wedge K_0$ does not depend on time, the retarded field 
(\ref{Fret3}) looks relatively simple:
\begin{eqnarray}\label{FretU}
{\hat F}^{\rm ret}(x)&=&e\int_{-\infty}^{\tau^{\rm ret}(x)}
\frac{{\rm d}\tau}{\sqrt{[\tau+(K_0\cdot u)]^2-(K_0\cdot u)^2-(K_0\cdot 
K_0)}}\frac{u\wedge K_0}{[-(K_0\cdot u)-\tau]^2}\nonumber\\
&=&\left.-e\frac{u\wedge K_0}{(K_0\cdot u)^2+(K_0\cdot K_0)}
\sqrt{1-\frac{(K_0\cdot u)^2+(K_0\cdot K_0)}{[-(K_0\cdot 
u)-\tau]^2}}\right|_{\tau\to-\infty}^{\tau=\tau^{\rm ret}(x)}\nonumber\\
&=&e\frac{u\wedge K_0}{(K_0\cdot u)^2+(K_0\cdot K_0)}.
\end{eqnarray}
Taking into account that the denominator is the square of the retarded 
distance (\ref{rint}) and substituting $u\wedge K$ for $u\wedge K_0$, we 
finally obtain
\begin{equation}\label{FrtU}
{\hat F}^{\rm ret}(x)=e\frac{u\wedge K}{r_{\rm ret}^2}.
\end{equation}
Can the uniformly moving charge ``fill'' this field? To answer this 
question we place the field point $x$ on particle's world line: 
$x^\mu=z_0^\mu+u^\mu\tau_x$. In such a case the separation vector 
$K^\mu=u^\mu(\tau_x-\tau)$ is collinear with particle's velocity. The 
strength tensor (\ref{FrtU}) vanishes on $\zeta$ and, therefore, the 
uniformly moving charge can not be accelerated due to its own field.

The simplest field is generated by a static charge placed at the 
coordinate origin. Setting $z=(t,0,0)$ and $u=(1,0,0)$ in 
eq.(\ref{FretU}), one can derive that the only nontrivial components of 
static field are:
\begin{eqnarray}\label{A1}
F_{i0}&=&e\int\limits_{-\infty}^{x^0-r}\frac{{\rm d}t}{\sqrt{(x^0-t)^2-r^2}}
\frac{x^i}{(x^0-t)^2}\\
&=&\left.-e\frac{x^i}{r^2}
\frac{\sqrt{(x^0-t)^2-r^2}}{x^0-t}\right|_{t\to -\infty}^{t=x^0-r}
=e\frac{x^i}{r^2}.\nonumber
\end{eqnarray}
$r:=\sqrt{(x^1)^2+(x^2)^2}$ is the distance to the charge. Since 
(\ref{Far2}), $F_{10}=E^1$ and $F_{20}=E^2$; magnetic component $H=0$. 

An interaction between two static charges, say $e_1$ and $e_2$, is caused 
by the Lorentz force $F^\alpha_{ab}=e_au_a^\beta F^\alpha{}_\beta$. 
Electric field (\ref{A1}) of the unmoved charge yields the stronger 
electrostatic force in comparison with its counterpart in conventional 
electrodynamics:
$$
{\mathbf F}_{ab}=e_1e_2\frac{{\mathbf n}_r}{r}.
$$
Unit vector 
${\mathbf n}_r:=({\mathbf z}_a-{\mathbf z}_b)/r, r=|{\mathbf z}_1-{\mathbf 
z}_2|$ is the distance between charges; ${\mathbf z}_a$ is position 
2-vector of $a$-th charge. The magnitude of the electrostatic force between 
two point charge is directly proportional to the magnitudes of each charge 
and inversely proportional to the distance between the charges.

Coulomb's law in three dimensions is similar to that in conventional 
spacetime of four dimensions. What is happen when the first clamp 
releases the charge? In four dimensions $e_1$ moves in electrostatic 
field of charge $e_2$. While in three dimensions the accelerated charge is 
influenced by its own field too! So, the ``static'' segment of the world 
line $\zeta_1$ generates the field
\begin{eqnarray}
F_{i0}&=&e_1\int\limits_{-\infty}^0\frac{{\rm d}t}{\sqrt{(x^0-t)^2-r^2}}
\frac{x^i}{(x^0-t)^2}\nonumber\\
&=&\left.-e_1\frac{x^i}{r^2}
\frac{\sqrt{(x^0-t)^2-r^2}}{x^0-t}\right|_{t\to -\infty}^{t=0}\nonumber\\
&=&e_1\frac{x^i}{r^2}\left(
1-\sqrt{1-\frac{r^2}{(x^0)^2}}
\right).\nonumber
\end{eqnarray}

Curvilinear segment of $\zeta_1$ (it corresponds to the time interval 
after initial instant $t=0$ when the charge becomes free) acts on $e_1$ too. 
The situation looks like a charge ``repulses'' itself taken in future 
instant of time. The ``self-force'' is inversely proportional to the 
distance between the ``momentary'' charge and the same charge taken at 
previous instant of time. What is happen when the points on the world line 
are very close?

To answer this question we evaluate the ``Lorentz self-force density''
\begin{eqnarray}\label{Lsf3}
f^\mu_{\rm ret}(\tau,s)&=&
\frac{e^2}{\sqrt{-(q\cdot q)}}\left\{
\frac{u_s^\mu(u_\tau\cdot q)-(u_\tau\cdot u_s)q^\mu}{r_s^2}\left[1+(q\cdot 
a_s)\right]\right.\\
&+&\left.\frac{a_s^\mu(u_\tau\cdot q)-(u_\tau\cdot a_s)q^\mu}{r_s}\right\},
\nonumber
\end{eqnarray}
in the $s\to\tau$ limit. (Expression (\ref{Lsf3}) is the contraction of the 
integrand of eq.(\ref{Fret3}) where field point $x=z(\tau)\in\zeta$, with 
particle's velocity $u(\tau)$.) Index $\tau$ indicates that the particle's 
velocity or position is referred to the actual instant $\tau$ while index 
$s$ says that the particle's characteristics are evaluated at previous 
instant $s<\tau$. Since both the field point $x=z(\tau)$ and the point of 
emission $z(s)$ lie on the same world line, the separation vector $K$ 
becomes $q(\tau,s)=z_\tau-z_s$; $r_s=-(q\cdot u_s)$. With a degree of 
accuracy sufficient for our purposes
\begin{eqnarray}
\sqrt{-(q\cdot q)}&=&\Delta,\nonumber\\
q^\mu&=&\Delta\left[u_\tau^\mu-a_\tau^\mu\frac{\Delta}{2}+
{\dot a}_\tau^\mu\frac{\Delta^2}{6}\right],\nonumber\\
u_s^\mu&=&u_\tau^\mu-a_\tau^\mu\Delta+{\dot 
a}_\tau^\mu\frac{\Delta^2}{2},\nonumber\\
a_s^\mu&=&a_\tau^\mu-{\dot a}_\tau^\mu\Delta,\nonumber
\end{eqnarray}
where $\Delta=\tau-s$ is positive small parameter. Substituting these into 
integrand of the double integral of eq. (\ref{Lsf3}) and passing to the 
limit $\Delta\to 0$ yields diverging expression:
\begin{equation}\label{lmt}
\lim_{s\to\tau}f^\mu_{\rm ret}(\tau,s)=
-\frac{e^2}{2}\lim_{\Delta\to 0}\frac{a^\mu_\tau}{\Delta} 
+\frac{2e^2}{3}\left({\dot a}_\tau^\mu-a_\tau^2u_\tau^\mu\right).
\end{equation}
Whence the ``Lorentz self-force'' $\int_{-\infty}^\tau f^\mu_{\rm 
ret}(\tau,s)$ can not be thought as the self-action in $2+1$ electrodynamics 
and the renormalization procedure is necessary.

\section{Regularization procedure}
\setcounter{equation}{0}

The main idea ought to be straightforward to formulate: the retarded field 
carries energy-momentum and angular momentum; outgoing waves remove energy, 
momentum, and angular momentum from the source which then undergoes a 
radiation reaction. Unfortunately, the problem is ambiguous since the field 
diverges in the neighbourhood of the particle's world line. A rule is 
necessary for extracting the appropriate finite parts of Noether quantities 
which exert the radiation reaction.

\begin{figure}[h]
\begin{center}
\epsfclipon
\epsfig{file=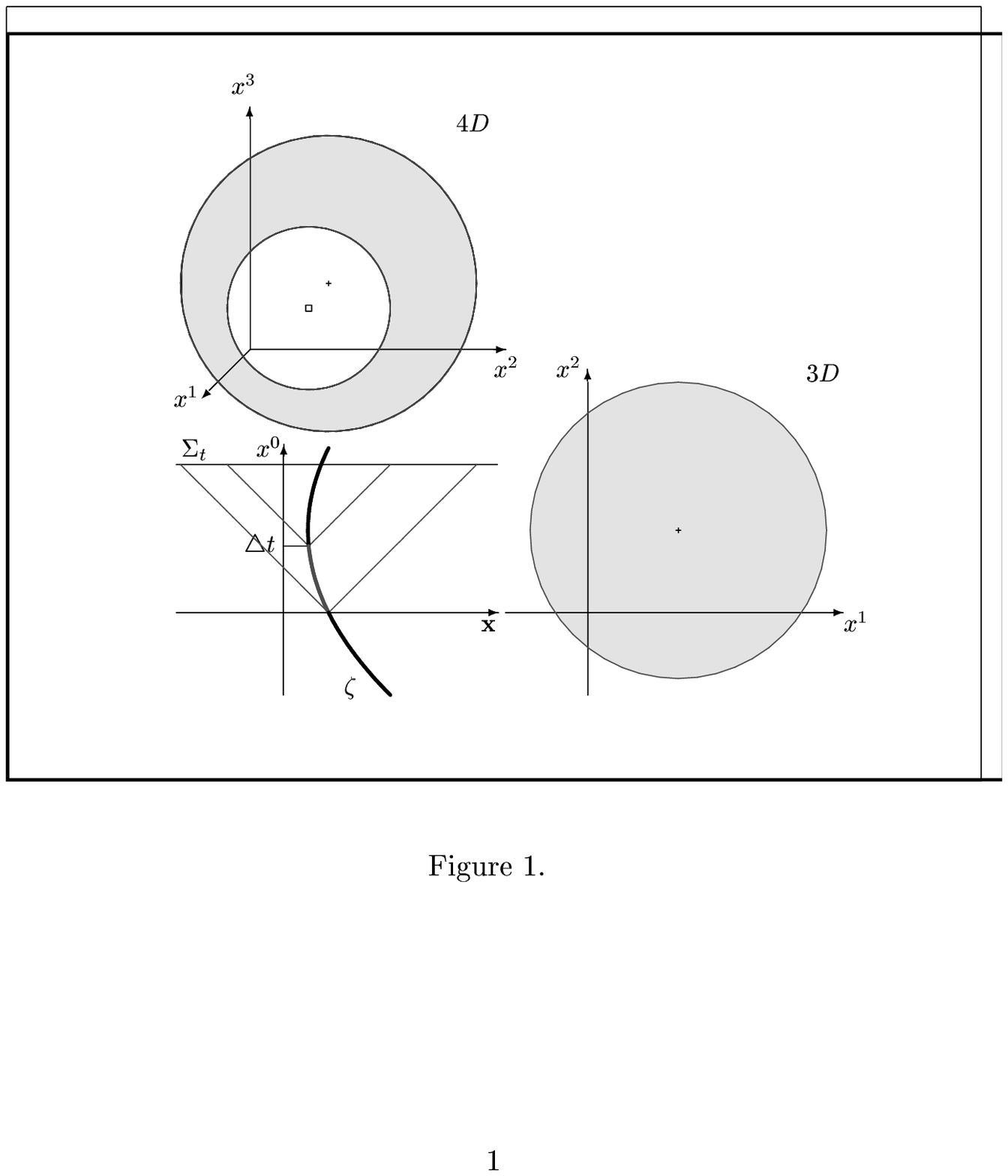,width=0.6\textwidth}
\end{center}
\caption{\label{Hg}
Let the point source radiates within the interval $[0,\triangle t]$. In four 
dimensions the support of the Maxwell energy-momentum tensor density in 
hyperplane $x^0=t$ is in between two spheres centered at points $z^i(0)$ 
(cross symbol) and $z^i(\triangle t)$ (box symbol) with radii $t$ and 
$t-\triangle t$, respectively. If the interval is infinitesimal, the support 
becomes two-dimensional sphere. In three dimensions the radiation fills the 
disk with radius $t$ centered at point $z^i(0)$ (cross symbol) even if the 
interval shrinks to zero.}
\end{figure}

The flow of energy-momentum that flows across a space-like surface $\Sigma$ 
is given by the integral \cite{Rohr}
\begin{equation}\label{pem3}
p^\nu_{\rm em}(\tau)=\int_\Sigma {\rm d}\sigma_\mu T^{\mu\nu},
\end{equation}
of the Maxwell energy-momentum tensor density
\begin{equation}\label{T3}
2\pi T^{\mu\nu}=F^{\mu\lambda}F^\nu{}_{\lambda}- 
1/4\eta^{\mu\nu}F^{\kappa\lambda}F_{\kappa\lambda}.
\end{equation}
over $\Sigma$. Since Huygens principle does not hold in three dimensions 
(see Figure \ref{Hg}) and field (\ref{Fret3}) develops a tail, the 
integration means the study of interference of outgoing electromagnetic 
waves emitted by different points on $\zeta$.

To calculate the flow of energy-momentum (\ref{pem3}), an appropriate 
surface of integration $\Sigma$ is necessary. The tilted hyperplane which 
plays privileged role in the one-particle radiation reaction problem in 
four dimensions \cite{Teit} is not suitable whenever the self-action 
problem in three dimensions is considered. The reason is that the amount of 
radiated energy-momentum in $\Sigma$ depends on all previous evolution of a 
source. There is no a plane which is orthogonal to the particle's 
3-velocities at {\it all} points on $\zeta$ before the end point 
$z(\tau)=\zeta\cap\Sigma$. We choose the simplest plane 
$\Sigma_t=\{x\in{\mathbb M}_{\,3}: x^0=t\}$ associated with an unmoving 
inertial observer. Non-covariant terms arise unavoidable due to integration 
over this surface. To reveal meaningful contribution in radiated 
energy-momentum we apply the criteria which were first formulated in 
\cite[Table 1]{Teit}:
\begin{itemize}
\item
the bound term diverges while the radiative one is finite;
\item
the bound component depends on the momentary state of the particle's
motion while the radiative one is accumulated with time; and
\item
the form of the bound terms heavily depends on choosing of an 
integration surface while the radiative terms are invariant. 
\end{itemize}

The second point should be define more accurately. In conventional 
electrodynamics the bound contribution (Shott term) depends on the momentary 
state of particle's motion while the radiated energy-momentum carried by 
electromagnetic field is the path integral of Larmor expression. In three 
dimensions both the bound term and the radiative one develop tails. But the 
radiative terms have one extra path integration in comparison with the bound 
ones.

The bound parts of Noether quantities modify particle's individual 
characteristics (its mass, its momentum and its angular momentum). 
To establish tail field contribution into particle's individual 
characteristics we do not manipulate with divergent non-covariant bound 
terms. We did not make any assumptions about the particle structure, its 
charge distribution, and its size. We only assume that the momentum of 
dressed charge is finite. To obtain additional information we calculate the 
flow of angular momentum \cite{Rohr}
\begin{equation}\label{M3}
M_{\rm em}^{\mu\nu}(t)=\int_{\Sigma_t} 
{\rm d}\sigma_0\left(x^\mu T^{0\nu} - x^\nu T^{0\mu}\right)
\end{equation}
which flows across $\Sigma_t=\{x\in{\mathbb M}_{\,3}: x^0=t\}$. To reveal 
radiative part of ${\hat M}_{\rm em}$ we apply Teitelboim's criteria. 
Further we assume that the bound terms are absorbed by particle's 
individual characteristics within the renormalization procedure while 
the radiative terms survive and lead an independent existence. The change in 
{\it radiative} energy-momentum and angular momentum carried by the 
electromagnetic field should be balanced by a corresponding change of 
particle's momentum and angular momentum, respectively. Analysis of six 
balance equations gives the form of individual characteristics of dressed 
charged particle as well as effective equation of motion which includes 
effect of particle's own field. 

\begin{figure}[ht]
\begin{center}
\epsfclipon
\epsfig{file=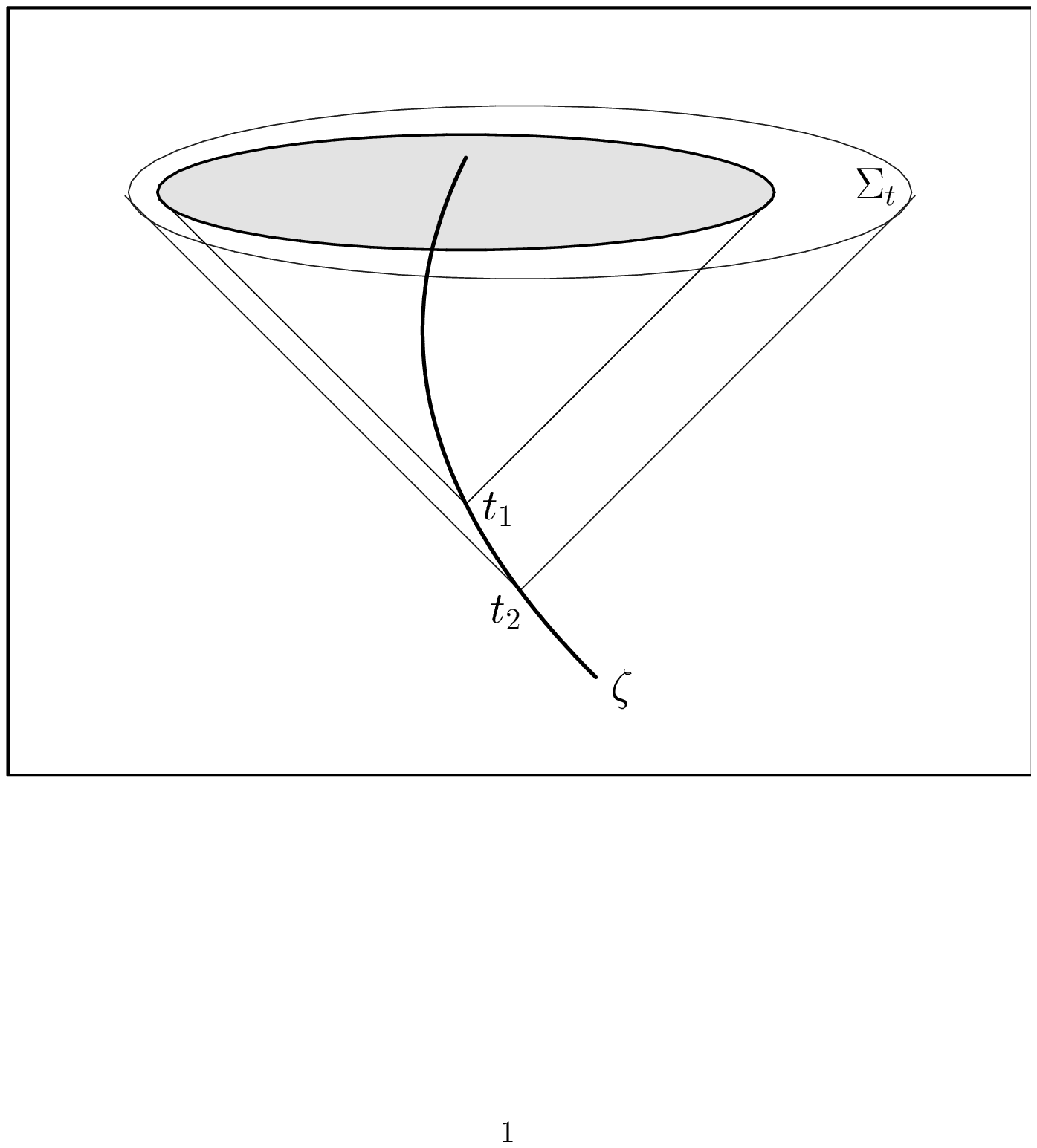,width=0.5\textwidth}
\end{center}
\caption{\label{int_ret}
Outgoing electromagnetic waves generated by the portion of the 
world line that corresponds to the interval $-\infty <t_2<t_1$ 
combine within the gray disk in the plane $\Sigma_t$. It is 
centered at point with coordinates $(z^1(t_1),z^2(t_1))$. Corresponding 
domain of integration is $\int_{-\infty}^\tau{\rm 
d}t_1\int_{-\infty}^{t_1}{\rm d}t_2
\int_0^{k_1^0}{\rm d}R\int_0^{2\pi}{\rm d}\varphi$ where $k_1^0$ is the 
radius of smaller disk.}
\end{figure} 

\subsection{Radiated energy-momentum of electromagnetic field}

Figure \ref{int_ret} pictures the interference of waves generated by 
fixed point $z(t_1)\in\zeta$ with radiation produced by charge before 
$t_1$. It is convenient to parametrize the plane $x^0=t$ by circular 
coordinates $R$ and $\varphi$. Since the stress-energy tensor 
(\ref{T3}) is quadratic in field strengths, we should {\it twice} 
integrate it over $\zeta$. The amount of radiated energy-momentum is 
given by fourfold integral
\begin{equation}\label{pnR}
p_{\rm R}^\nu=\int\limits_{-\infty}^t{\rm 
d}t_1\int\limits_{-\infty}^{t_1}{\rm d}t_2
\int\limits_{0}^{k_1^0}{\rm d}R \int\limits_0^{2\pi}{\rm d}\varphi 
Jt^{0\nu}_{12},
\end{equation}
where $J$ is Jakobian of coordinate transformation \cite[eq.(4.5)]{Yar3D}. 
The integrand
\begin{equation}\label{t12}
2\pi t^{\alpha\beta}_{12}=
f_{(1)}^{\alpha\lambda}f_{(2)\lambda}^\beta -\frac14\eta^{\alpha\beta}
f_{(1)}^{\mu\nu}f^{(2)}_{\mu\nu}
\end{equation}
describes the combination of field strength densities at $x\in\Sigma_t$
$$
f_{\mu\nu}=\frac{e}{\sqrt{-(K\cdot K)}}\left\{
\frac{u_\mu K_\nu - u_\nu K_\mu}{[-(K\cdot u)]^2}\left[1+(K\cdot a)\right]
+\frac{a_\mu K_\nu - a_\nu K_\mu}{-(K\cdot u)}
\right\},
$$
generated by emission points $z(t_1)\in\zeta$ and $z(t_2)\in\zeta$.

The calculation is performed in Ref.\cite{Yar3D} where radiative terms are 
extracted. Resulting expressions can be rewritten in a manifestly covariant 
fashion when the world line is parameterized by a proper time $\tau$.
The fourfold integral (\ref{pnR}) contributes in {\it radiated} 
energy-momentum one-half of the work 
$$
p_{\rm R}^\mu=-\frac{1}{2}  
\int_{-\infty}^\tau{\rm d}\tau_1 F_{\rm ret}^\mu(\tau_1),
$$
of the retarded tail Lorentz force
\begin{eqnarray}\label{LFret}
F_{\rm ret}^\mu(\tau_1)&=& eF_{(\theta)}^{\mu\alpha}u_{1,\alpha}\\
&=&e^2
\int_{-\infty}^{\tau_1}{\rm d}\tau_2\frac{-(q\cdot u_1)u_2^\mu 
+(u_1\cdot u_2)q^\mu}{[-(q\cdot q)]^{3/2}},\nonumber
\end{eqnarray}
taken with opposite sign.

\begin{figure}[ht]
\begin{center}
\epsfclipon
\epsfig{file=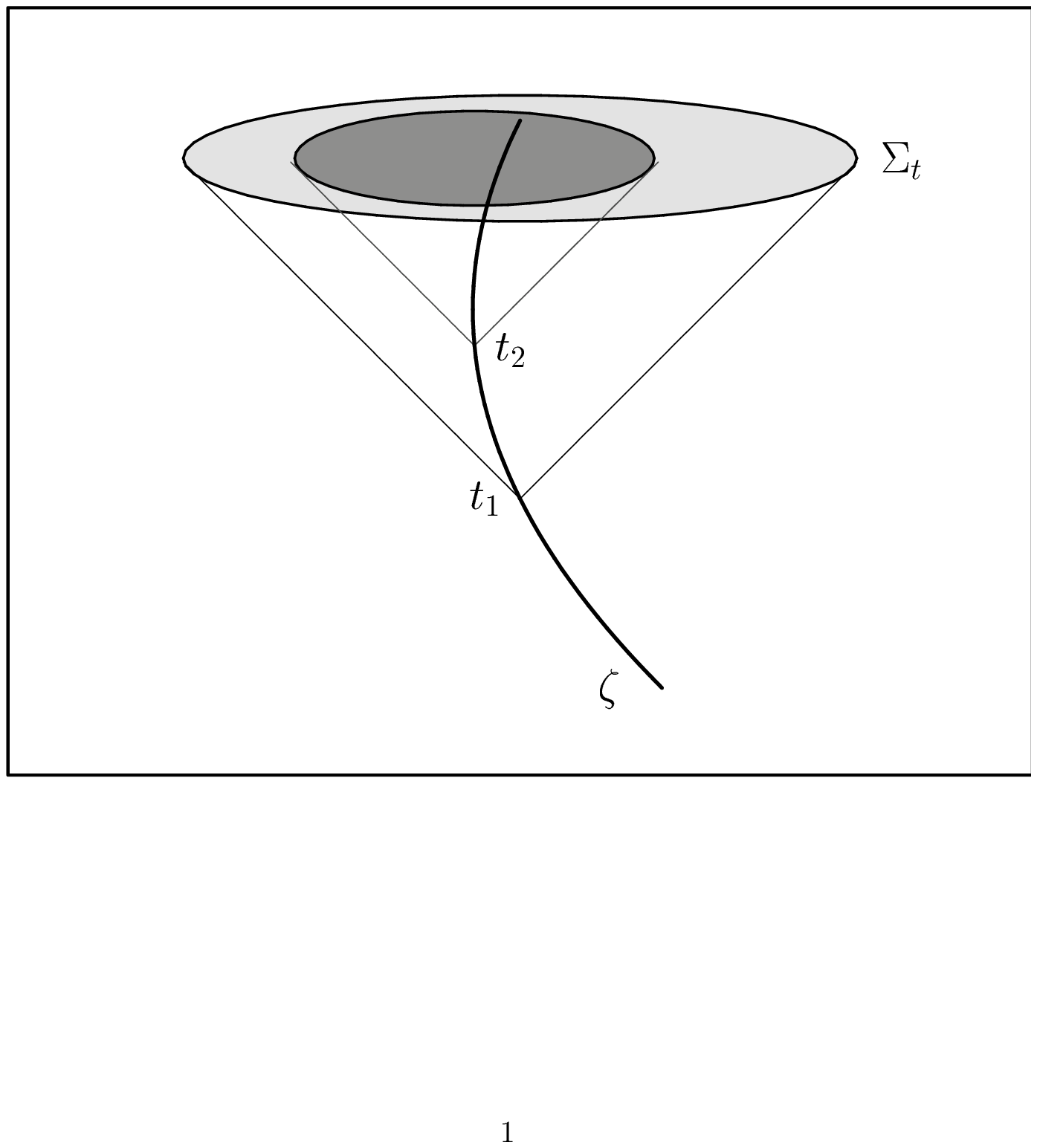,width=0.45\textwidth}
\end{center}
\caption{\label{int_adv}
Outgoing electromagnetic waves generated by the portion of the 
world line that corresponds to the interval $t_1<t_2\le t$  
joint together inside the dark disk. It is centered at point with 
coordinates $(z^1(t_2),z^2(t_2))$. The domain of integration becomes 
$\int_{-\infty}^t{\rm d}t_1\int^{t}_{t_1}{\rm d}t_2
\int_0^{k_2^0}{\rm d}R\int_0^{2\pi}{\rm d}\varphi$ where $k_2^0$ is the 
radius of dark disk.}
\end{figure} 

Figure \ref{int_adv} illustrates the interference of the radiation 
emanated by fixed point $z(t_1)\in\zeta$ with waves generated by charge 
during the interval $[t_1,t]$. The portion of energy-momentum produced 
by this segment of $\zeta$ is given by the multiple integral
$$
p_{\rm A}^\nu=\int\limits_{-\infty}^t{\rm 
d}t_1\int\limits^t_{t_1}{\rm d}t_2
\int\limits_{0}^{k_2^0}{\rm d}R \int\limits_0^{2\pi}{\rm d}\varphi 
Jt^{0\nu}_{12},
$$
where tensor $t^{0\nu}_{12}$ is defined by eq.(\ref{t12}). This fourfold 
integral contributes in {\it radiated} energy-momentum one-half of work 
$$
p_{\rm A}^\mu=-\frac{1}{2}  
\int_{-\infty}^\tau{\rm d}\tau_1 F_{\rm adv}^\mu(\tau_1),
$$
of the advanced tail Lorentz force
\begin{equation}\label{LFadv}
F_{\rm adv}^\mu(\tau_1)=e^2
\int^\tau_{\tau_1}{\rm d}\tau_2\frac{-(q\cdot u_1)u_2^\mu 
+(u_1\cdot u_2)q^\mu}{[-(q\cdot q)]^{3/2}},
\end{equation}
that differs from its retarded counterpart (\ref{LFret}) by the domain of 
integration only. 

\begin{figure}[t]
\begin{center}
\epsfclipon
\epsfig{file=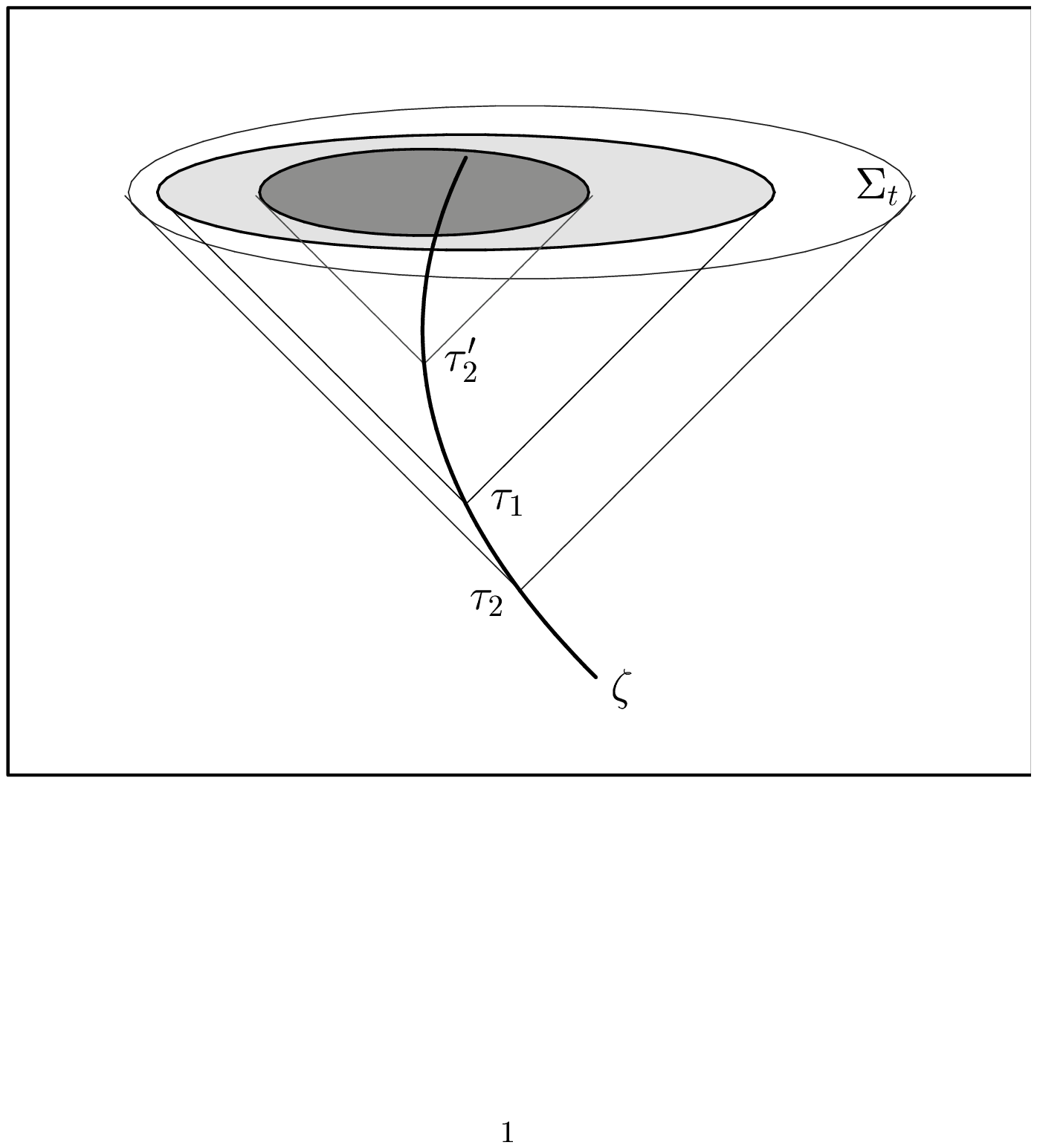,width=0.45\textwidth}
\end{center}
\caption{\label{int_ra}
We call ``retarded'' the force (\ref{LFret}) with integration over the 
portion of the world line {\it before} $\tau_1$. We call ``advanced'' the 
force (\ref{LFadv}) with integration over the portion of the world line 
{\it after} $\tau_1$. For an observer placed at point $z(\tau_1)\in\zeta$ 
the regular radiative part (\ref{prad3}) of electromagnetic field momentum 
looks as the combination of incoming and outgoing radiation. And yet the 
retarded causality is not violated. We still consider the interference of 
outgoing waves presented at the observation instant $\tau$. The 
electromagnetic field carries information about the charge's past.}
\end{figure}

The radiative part of energy-momentum carried by electromagnetic field is 
therefore
\begin{equation}\label{prad3}
p_{\rm rad}^\mu(\tau)=-\frac12\int_{-\infty}^\tau {\rm d}\tau_1\left( 
F_{\rm ret}^\mu-F_{\rm adv}^\mu\right).
\end{equation}
The resulting expression obeys the spirit of Dirac's decomposition of 
the retarded electromagnetic field into the ``mean of the advanced and 
retarded field'' and the ``radiation'' field. The situation is pictured in 
Figure \ref{int_ra}.

\subsection{Radiated angular momentum of electromagnetic field}
\setcounter{equation}{0}

Surface integration of the torque of the stress-energy tensor (\ref{M3}) is 
identical to that of the energy and momentum tensor densities. Details of 
computations is presented in ref.\cite{Yar3D}. The amount of radiated 
angular momentum that corresponds to interference pictured in figure 
\ref{int_ret} is as follows:
$$
M_{\rm ret}^{\mu\nu}=-\frac{1}{2}  
\int_{-\infty}^\tau{\rm d}\tau_1 \left[
z_1^\mu F_{\rm ret}^\nu(\tau_1)-z_1^\nu F_{\rm ret}^\mu(\tau_1)
\right].
$$
It is then nothing but one-half of the path integral of the torque of 
tail Lorentz force (\ref{LFret}). Combination of waves pictured in figure 
\ref{int_adv} contributes in radiated angular momentum one-half of the path 
integral of the torque of ``advanced'' Lorentz force (\ref{LFadv}):
$$
M_{\rm adv}^{\mu\nu}=-\frac{1}{2}  
\int_{-\infty}^\tau{\rm d}\tau_1 \left[
z_1^\mu F_{\rm adv}^\nu(\tau_1)-z_1^\nu F_{\rm adv}^\mu(\tau_1)
\right].
$$
Total angular momentum which leads an independent existence and can be 
detected by distant devices is as follows:
\begin{eqnarray}\label{Mrad3}
M_{\rm rad}^{\mu\nu}(\tau)&=&
-\frac12\int_{-\infty}^\tau {\rm d}\tau_1\left( 
M_{\rm ret}^{\mu\nu}-M_{\rm adv}^{\mu\nu}\right)\\
&=&-\frac12\int_{-\infty}^\tau {\rm 
d}\tau_1\left[ z_1^\mu\left(F_{\rm ret}^\nu-F_{\rm adv}^\nu\right)-
z_1^\nu\left(F_{\rm ret}^\mu-F_{\rm adv}^\mu\right)\right].\nonumber
\end{eqnarray}
Together with radiated energy-momentum (\ref{prad3}), it exert the 
radiation reaction. 

\section{Equation of motion of radiating charge}\label{beq}
\setcounter{equation}{0}

We therefore introduce the radiative part (\ref{prad3}) of energy-momentum 
and postulate that it, and it alone, exerts a force on the particle. 
Singular part should be coupled with particle's three-momentum, so that 
``dressed'' charged particle would not undergo any additional radiation 
reaction. Already renormalized particle's individual three-momentum, say 
$p_{\rm part}$, together with $p_{\rm rad}$ constitute the total 
energy-momentum of our composite particle plus field system:  $P=p_{\rm 
part}+p_{\rm rad}$.

The total angular momentum, say $\hat M$, consists of particle's 
angular momentum $z\wedge p_{\rm part}$ and radiative part (\ref{Mrad3}) of 
angular momentum carried by electromagnetic field:
$$
M^{\mu\nu}=z_\tau^\mu p_{\rm part}^\nu(\tau) 
- z_\tau^\nu p_{\rm part}^\mu(\tau) + M^{\mu\nu}_{\rm rad}(\tau).
$$

The one-half sum of the retarded and the advanced works is the bound 
part of tail energy-momentum which is permanently attached to the charge 
and is carried along with it. It modifies particle's individual
characteristics (its momentum and its inertial mass). A point source 
together with surrounded electromagnetic ``cloud'' constitute new entity: 
dressed charged particle.

Balance equations ${\dot P}=F_{\rm ext}$ and ${\dot M}=z_\tau\wedge 
F_{\rm ext}$ result integro-differential equation of motion of a dressed 
charged particle in an external field
\begin{equation}\label{mext}
ma^\mu_\tau=eu_{\tau,\alpha} F^{\mu\alpha}_{\rm ret}(\tau)+ 
\frac{e^2}{2}a^\mu_\tau\int_{-\infty}^\tau 
\frac{{\rm d} s}{\sqrt{-(q\cdot q)}}
+eu_{\tau,\alpha} F^{\mu\alpha}_{\rm ext},
\end{equation} 
where the radiation reaction is taken into account. The non-local term 
in equation (\ref{mext}) which is proportional to particle's acceleration 
$a(\tau)$ arises also in \cite{KLS}. It provides proper short-distance 
behavior of the perturbations due to the particle's own field. If $s\to\tau$ 
the integrand tends to three-dimensional analog of the Abraham radiation 
reaction vector:
$$
\lim_{s\to\tau}\left[
eu_{\tau,\alpha}f^{\mu\alpha}(\tau,s)+
\frac{e^2}{2}\frac{a^\mu_\tau}{\sqrt{-(q\cdot q)}}
\right]=\frac23e^2\left({\dot a}^\mu-a^2u^\mu\right),
$$
cf. eq.(\ref{lmt}). All quantities on the right-hand side refer to the 
instant of observation $\tau$.

Individual 3-momentum of a dressed charged particle contains nonlocal 
contribution from tail electromagnetic field of the particle:
$$
p_{\rm part}^\mu(\tau)=mu^\mu(\tau)+\frac{e^2}{2}\int_{-\infty}^\tau 
{\rm d} s\frac{u^\mu(s)-u^\mu(\tau)}{\sqrt{-(q\cdot q)}}.
$$
The balance equations produces also a time-changing inertial mass:
\begin{equation}\label{dotm}
\dot m=\frac{e^2}{2}\int_{-\infty}^\tau {\rm d}s
\frac{(q\cdot u_\tau)-(q\cdot u_s)}{[-(q\cdot q)]^{3/2}}.
\end{equation}
It is interesting that similar phenomenon occurs in the theory which 
describes a point-like charge coupled with massless scalar field in flat 
spacetime of three dimensions \cite{Br}. The charge loses its mass through 
the emission of monopole radiation.

\section{Radiating charge in uniform static electric field}
\setcounter{equation}{0}

Let us consider a constant electromagnetic field with components
$$
\left(F^{\mu\nu}_{\rm ext}\right)=
\left(
\begin{array}{ccc}
0 & E & 0\\
-E& 0 & 0\\
0 & 0 & 0
\end{array}
\right)
$$
acting on the charge $e$ during the interval $[0,\tau_0]$. Before the 
initial instant $\tau=0$ the particle places at the coordinate origin.

If we neglect the self-action, the equation (\ref{mext}) becomes
$$
\frac{{\rm d}u^0}{{\rm d}\tau}=\frac{e}{m}Eu^1,\qquad
\frac{{\rm d}u^1}{{\rm d}\tau}=\frac{e}{m}Eu^0,\qquad
\frac{{\rm d}u^2}{{\rm d}\tau}=0.
$$
The test charge moves along hyperbola
\begin{equation}\label{wl}
z^0(\tau)=a^{-1}\sinh(a\tau),\qquad
z^1(\tau)=a^{-1}\left[\cosh(a\tau)-1\right],\qquad
z^2(\tau)=0,
\end{equation}
where constant
$$
a=\frac{e}{m}E
$$
is the modulo of acceleration:
\begin{eqnarray}\label{ua}
u^0(\tau)&=&\cosh(a\tau),\quad z^1(\tau)=\sinh(a\tau),\quad u^2(\tau)=0,\\
a^0(\tau)&=&a\sinh(a\tau),\quad z^1(\tau)=a\cosh(a\tau),\quad 
a^2(\tau)=0.\nonumber
\end{eqnarray}

Let us suppose that external field is much more than electrostatic field 
generated by static segment of the world line, so that the radiation 
reaction does not change the type of particle's world line. We substitute 
the solution (\ref{wl}) and its differential consequences (\ref{ua}) for 
corresponding quantities in the self-action terms of the equation of motion 
of radiating charge (\ref{mext}). So, the distance between charge at 
instant of observation $\tau<\tau_0$ and the same charge taken at moment 
$0<s<\tau$ depends on the difference $\tau-s$
$$
\sqrt{-(q\cdot q)}=\frac2a\sinh[\frac{a}{2}(\tau-s)]. 
$$
Hence the derivative of dynamical mass (\ref{dotm}) vanishes.

It is of great importance that radiation back reaction is proportional to 
particle's acceleration:
\begin{eqnarray}
e^2\int_0^\tau{\rm d}s\left[
u_\alpha(\tau)\frac{f^{\mu\alpha}(\tau,s)}{\sqrt{-(q\cdot q)}}+
\frac12\frac{a^\mu(\tau)}{\sqrt{-(q\cdot q)}}
\right]&=&
\left.\frac{e^2}{2}\frac{a^\mu(\tau)}{\cosh[a(\tau-s)/2]}\right|_{s=0}^{s=\tau}\nonumber\\
&=&\frac{e^2}{2}a^\mu(\tau)\left[
1-\frac{1}{\cosh(a\tau/2)}
\right].\nonumber
\end{eqnarray} 
The portion of trajectory where the second term in between the square 
brackets is much less than $1$ approximates to hyperbola like (\ref{wl})
where modified mass $m-e^2/2$ should be substituted for $m$. The solution 
of the equation of motion (\ref{mext}) is given by functions (\ref{wl}) 
where new modulo of acceleration
$$
a=\frac{eE}{m-e^2/2}
$$
is greater than initial one. It is because the Lorentz self-force is 
repulsive (the charge interacts with itself).

After interaction zone, the charge fills the effect of radiation emitted 
by curvilinear portion of the world line. In general, the particle do not 
move uniformly nevermore. However, the contributions to the tail terms 
arising from portions of the trajectory distant to the current position of 
the particle should become negligible and its velocity will tend to a 
constant value.

\section{Conclusions}
\setcounter{equation}{0}

In the present paper, we adopt the Dirac scheme of decomposition of the 
retarded Green's function into symmetric (singular) and radiative (regular) 
parts to functions supported within light cones. The regularization scheme 
summarizes a scrupulous analysis of energy-momentum and angular momentum 
balance equations in $2+1$ electrodynamics \cite{Yar3D,Y3D}. It differs 
from the approach developed by Detweiler and Whiting \cite{DW} on two 
``extra'' entities: additional instant $\tau_1$ before the instant of 
observation and extra integration of the ``half-difference'' of the retarded 
and the advanced tail forces over particle's path. So, the retarded tail 
force depends on the particle's past history before $\tau_1<\tau$. Its 
advanced counterpart is generated by portion of the world line that 
corresponds to the interval $[\tau_1,\tau]$. The tail part of radiated 
energy-momentum is one-half of the work done by the retarded force minus 
one-half of the work done by the advanced force, taken with opposite sign. 
This part of radiation detaches the point source and leads an independent 
existence.

The one-half sum of the retarded and the advanced works is the bound 
part of tail energy-momentum which is permanently attached to the charge 
and is carried along with it. It modifies particle's individual
characteristics (its momentum and its inertial mass). 

Since the properties of the retarded and the advanced solutions of wave 
equation, the one-half sum is singular while one-half difference is regular
in the immediate vicinity of the world line.

The items can be summarized as a simple scheme which obeys the spirit of 
Dirac's scheme of decomposition of the retarded field in conventional 
electrodynamics into singular and regular parts. The main points are as 
follows.
\begin{itemize}
\item
The tail retarded field can be decomposed into symmetric (singular) and 
radiative (regular) parts in standard Dirac's manner.
\item
The support of both the retarded field and the advanced field is limited to 
particle's world line. 
\end{itemize}

The bound and the radiative angular momentum carried by charge's  
field are simply torques of the above combinations of the retarded and the 
advanced tail forces.

Changes in individual momentum and angular momentum of dressed charged 
particle compensate losses of energy, momentum, and angular momentum due 
to radiation. (Influence of an external device can be modelled easily.) 
Analysis of balance equations gives the self-force and three-dimensional 
analogue of the Lorentz-Dirac equation.

\section*{Acknowledgments}
I am grateful to V.Tre\-tyak for continuous encouragement and for 
a helpful reading of this manuscript. I would like to thank A.Du\-vi\-ryak 
and O.Derzhko for useful discussions.

\end{document}